\documentclass[aps,pra,amssymb,amsfonts,floatfix,showpacs]{revtex4}

\usepackage{graphicx}
\usepackage{dcolumn}
\usepackage{bm}
\usepackage{mathbbol}

\begin{document}

\title{Interplay between interaction and (un)correlated disorder in 
one-dimensional many-particle systems: delocalization and global entanglement}

\author{Frieda Dukesz}

\author{Marina Zilbergerts}

\author{Lea F. Santos} \thanks{Corresponding author: {\tt lsantos2@yu.edu}}
\affiliation{\mbox{Department of Physics, Yeshiva University, 245 Lexington Ave, New York, NY, 10016, USA}}

\date{\today}

\begin{abstract}
We consider a one-dimensional quantum many-body system and investigate how 
the interplay between  interaction and on-site disorder affects spatial localization and quantum correlations. The hopping amplitude is kept constant.
To measure localization, we use the number of principal components (NPC),
which quantifies the spreading of the system eigenstates over vectors of a given basis set.
Quantum correlations are determined by a global entanglement measure $Q$, which quantifies the
degree of entanglement of multipartite pure states.
Our studies apply analogously to a one-dimensional system of
interacting spinless fermions, hard-core bosons, or yet
to an XXZ Heisenberg spin-1/2 chain. 
Disorder is characterized by both: uncorrelated and long-range correlated random on-site energies. 
Dilute and half-filled chains are analyzed. 
In half-filled clean chains, delocalization is maximum when the particles do not interact, 
whereas multi-partite entanglement is largest when they do.
In the presence of uncorrelated disorder, NPC and Q show a non-trivial behavior
with interaction, peaking in the chaotic region.
The inclusion of correlated disorder may further extend two-particle states, but the effect decreases with the number of particles and strength of their interactions. 
In half-filled chains with large interaction,
correlated disorder may even enhance localization.
\end{abstract}

\date{\today}
\maketitle

\section{Introduction}

Disorder may significantly affect the properties of physical systems.
Spatial localization of one-particle states (Anderson localization), for example, 
is due to uncorrelated random disorder~\cite{Anderson1958,Lee1985,LifshitzBook,Kramer1993}; whereas
short range~\cite{Flores1989,Dunlap1990,Sanchez1994} and long range correlated 
\cite{Moura1998,Izrailev1999} disorder
have been associated with the appearance of delocalized states. Interestingly, however,
a recent experiment~\cite{Kuhl2008} shows that long-range correlated disorder may 
in fact either suppress or enhance localization.
This scenario becomes yet more complex when two or more particles are 
considered and the effects of interactions are taken into account.
A clear picture of the interplay between interaction and disorder
is essential to advance our understanding of 
thermodynamic and kinetic properties, as well as the dynamical
behavior of real quantum many-body systems. 

The interest in how interactions influence localization and transport properties of mesoscopic systems
was boosted with the realization that persistent currents could
only be explained if the role of the electron-electron interaction
was addressed \cite{Guhr1998}. 
In general, interaction between particles in disordered systems may lead to delocalization 
\cite{Chui1977,Apel1982,Kagan1984,Giamarchi1988,Dorokhov1990,Shepelyansky1994,Evangelou1996,
Abrahams2001,Nascimento2005,Dias2007}.
The competition between 
interaction and disorder in many-body systems relates also to other
interesting phenomena, such as the rich variety of quantum phase transitions of
ultracold atomic Bose gases in optical lattices 
\cite{Roth2003},
the transition from integrability to chaos \cite{Georgeot1998,Avishai2002,Santos2004}
in spin systems,
and the enhancement of entanglement 
\cite{SantosEscobar2004,Lakshminarayan2005,Monasterio2005b,Brown2008} 
and the `melting' of quantum computers \cite{Georgeot2000} in the context of
quantum information.

In the present work we investigate how disorder combined with 
interactions affect spatial localization and quantum correlations of
a one-dimensional quantum system. In addition to uncorrelated
disorder, we analyze also long-range correlated 
disorder. Correlated disorder may appear in real systems~\cite{Flambaum1999}
and may also be engineered. The latter includes the introduction 
of scatterers~\cite{Kuhl2008} or speckles~\cite{Billy2008} 
in a one-dimensional waveguide
or yet the individual tuning 
of on-site energies via local 
fields~\cite{Dykman2000,Platzman2000,Santos2005loc}.
In contrast to the extensively studied Hubbard model, where 
interaction occurs between particles in the same site,
we consider interaction between particles in neighboring sites.
Both dilute and half-filled chains
are analyzed. We study the level of delocalization,
determined by the number of principal components NPC,
and the amount of multi-partite entanglement,
quantified by a global entanglement measure $Q$, of all 
eigenvectors of the system. 
Our numerical results show that: (i) In a
non-interacting system, 
NPC and $Q$ decrease with uncorrelated on-site disorder and 
increase with correlated on-site disorder; 
(ii) In a clean system, interaction restricts delocalized states 
to narrow energy bands and the average
spatial delocalization decays,
while the behavior of quantum correlations is non-monotonic; 
(iii) In a disordered
system the behavior 
of NPC and $Q$ with interaction depends highly on the number of particles,
being non-trivial in a half-filled chain, where quantum chaos may develop;
(iv) On average, 
states of two-interacting particles appear to be delocalized
in the considered finite systems with uncorrelated disorder, 
while one-particle states localize.

The paper is organized as follows. Sec.~II describes the model,
the relation determining on-site disorder,
and the quantities computed. The half-filled chain is studied
in Sec.~III, disordered systems with non-interacting, weakly interacting and
strongly interacting particles are compared.
Sec.~IV considers the dilute limit and compares the results
for one and two-particle states.
Discussions and concluding remarks are presented in Sec.~V.

%

\section{System Model and Measures of Delocalization and Quantum Correlations}

\subsection{System Model}

The analysis developed here applies to different one-dimensional quantum many-body systems, 
including XXZ Heisenberg spin-1/2 models and chains of interacting spinless fermions or
 hard-core bosons. 

The Heisenberg spin-1/2 chain 
describes quasi-one-dimensional magnetic 
compounds~\cite{Salunke2007} and
Josephson-junction-arrays~\cite{Glazman1997,Giuliano2005}. 
It has also been broadly used as a model for 
proposals of quantum computers, 
including those based on semiconductor quantum dot 
arrays~\cite{Loss1998}, solid state NMR spin systems~\cite{Baugh2006}, and electrons floating on Helium~\cite{Platzman2000,Dykman2000}. 
We investigate a chain with
open boundary conditions and nearest-neighbor interactions, as determined by the Hamiltonian:

\begin{eqnarray}
&&H = H_{0i} + H_{\text{XY}}=H_0 + H_{\text{int}}+ H_{\text{XY}} ,\nonumber \\
&&H_0=\sum_{n=1}^{L} \Omega_n S_n^z,\nonumber \\
&&H_{\text{int}} \sum_{n=1}^{L-1} J \Delta S_n^z S_{n+1}^z ,\nonumber \\
&&H_{\text{XY}}= \sum_{n=1}^{L-1} J \left( 
 S_n^x S_{n+1}^x + S_n^y S_{n+1}^y  \right)\:.
\label{ham}
\end{eqnarray}
Above,  $\hbar$ is set equal to 1,
$L$ is the number of sites, and $\vec{S}_n = \vec{\sigma}_n/2$ is the spin operator at site $n$,   $\sigma^{x,y,z}_n$ being the Pauli operators. 
The parameter $\Omega_n= \omega + \omega_n$, 
where $\omega_n= d \epsilon_n$, corresponds to the Zeeman splitting of spin $n$, 
as determined by a static magnetic field in the $z$ direction.
In a clean system, all sites have the same energy splitting ($d=0$), whereas
disorder is characterized by the presence of on-site defects ($d\neq0$). 
The relation specifying $\epsilon_n$ is discussed in Sec.~II.B --
correlated and uncorrelated random disorder are considered. 
$J$ is the exchange coupling strength and $\Delta $ is the anisotropy associated with the
Ising interaction $S_n^z S_{n+1}^z$. We set $J, \Delta>0$.
The total spin operator in the
$z$ direction, $S^z=\sum_{n=1}^L S_n^z$, is conserved, therefore
the matrix $H$ is composed of independent blocks. Each block belongs to a
single $S^z$ subspace characterized by a fixed number $M$ of total spins pointing up,
the dimension $N$ of each sector is given by $N={L \choose M}=L!/[(L-M)!M!]$.

The model of Eq.~\ref{ham} may be mapped into a spinless fermion system via 
a Jordan-Wigner transformation \cite{Jordan1928}, so that
$H_{\text{XY}}$ becomes $H_{\text{hop}}$ and

\begin{eqnarray}
&&H =  H_{0i} + H_{\text{hop}}=H_0 + H_{\text{int}}+ H_{\text{hop}} ,\nonumber \\
&&H_0 = \sum_{n=1}^{L} \Omega_n  a_n^{\dagger} a_n \nonumber \\
&&H_{\text{int}}= 
\sum_{n=1}^{L-1} J \Delta a_n^{\dagger} a_{n+1}^{\dagger} a_{n+1} a_{n},
\nonumber \\
&&H_{\text{hop}}= \sum_{n=1}^{L-1}  
\frac{J}{2} (a_n^{\dagger} a_{n+1} + a_{n+1}^{\dagger} a_{n})\:,
\label{ham_fermions}
\end{eqnarray}
Above, $a_{n}^{\dagger}$ and $a_n$ are creation and annihilation operators, respectively.
The presence of a fermion on site $n$ corresponds to an excited spin, the on-site
fermion energies are the Zeeman energies, $J$ is the fermion hopping integral, and $J\Delta$
gives the fermion interaction strength. The conservation of $S^z$ translates here into
conservation of the total number of particles.
Disordered systems of interacting fermions simplified by neglecting the spin
degrees of freedom has been vastly considered as a first
approximation in studies of the metal-insulator transition~\cite{Giamarchi1988,Uhrig1993}.

The fermionic model above may also be mapped onto a system of hard-core bosons.
The Hamiltonian has the same form of Eq.~\ref{ham_fermions}, where the
fermionic creation and annihilation operators are substituted 
by bosonic ones~\cite{Rigol2007}. A system of hard-core bosons in one-dimensional optical
lattices constitutes a versatile
tool in studies of various complex quantum-physical phenomena, 
and has been receiving enormous attention, specially after experimental 
realizations~\cite{Paredes2004}.

Here, a particle or an excitation will generically refer to an
excited spin, a spinless fermion, or a
hard-core boson, depending on the specific system one addresses. 
We assume $L$ even and study both  
a half-filled ($M=L/2$) and a dilute ($M=2$) chain.

\subsection{On-site disorder}

There are widely different physical, biological, and economical
processes modeled as stochastic time series. These series
may be generated, for example, by imposing a power law power 
spectrum $S(f) \propto f^{-\alpha}$\cite{Osborne1989,Greis1991}.
The value of $\alpha$ determines the type of noise:
$\alpha =0, 1$ and 2 correspond, respectively, to white noise
(flat frequency spectrum), 1/f noise, and Brownian noise.
We borrow from these studies the specific sequence of long-range correlated
on-site energies considered in this work,

\begin{equation}
\epsilon_n = \sum_{k=1}^{L/2} \left[ \sqrt{k^{-\alpha} 
\left| \frac{2\pi}{L} \right|^{1-\alpha}} 
\cos \left( \frac{2\pi n k}{L} + \phi_k\right) \right],
\label{long-range}
\end{equation}
where $\phi_k$ are random numbers uniformly distributed in the range $[0,2\pi]$.
The sequence is constructed so that, by Fourier transforming
the two-point correlation function $\langle \epsilon_n \epsilon_m \rangle$,
one obtains the power law spectral density 
$S(k) \propto k^{-\alpha}$.
When $\alpha=0$, $\epsilon_n$'s in $\omega_n= d \epsilon_n$ are
random numbers with a Gaussian distribution, 
leading to the scenario of uncorrelated disorder: 
$\langle \omega_n \rangle=0$
and $\langle \omega_n \omega_m \rangle =d^2 \delta_{n,m}$.
Correlated on-site energies appear for
$\alpha>0$. 
The energy sequence is normalized: $\langle \epsilon_n \rangle = 0$
and the unbiased dispersion
$\sqrt{ \sum_{n=1}^{L-1} (\epsilon_n - \langle \epsilon_n \rangle)^2/(L-1)} =1$.

Long-range correlations are widespread in biological physics and
have been extensively analyzed in this context \cite{Stanley1993}.
In the field of condensed-matter physics, 
most works dealing with 
short-range~\cite{Flores1989,Dunlap1990,Sanchez1994} and
long-range~\cite{Moura1998,Izrailev1999}
correlated disorder in the context
of mobility edges have been limited to the 
case of a single particle. Ref.~\cite{Moura1998}, for instance, 
considered sequence (\ref{long-range}) and showed that when $\alpha>2$ 
the one-particle
wave functions remain delocalized even in the thermodynamic limit.
Here, two or more excitations are taken into account
and we investigate how the disorder parameters $d$ and $\alpha$
and the interaction amplitude $\Delta$ affect
delocalization and multipartite entanglement in the finite systems 
described above.

\subsection{Delocalization}

To quantify the extent of 
delocalization of an eigenvector 
$|\psi_j \rangle = \sum_{k=1}^N c^k_j |\varphi^k \rangle $
of Hamiltonian (\ref{ham},\ref{ham_fermions}) written in the basis
$|\varphi^k \rangle$, we consider the number of principal 
components ($\mbox{NPC}$)
\cite{Izrailev1990}, defined as

\begin{equation}
\mbox{NPC}_j \equiv \frac{1}{\sum_{k=1}^{N} |c^k_j|^4} .
\label{NPC}
\end{equation}
This quantity is also commonly referred to as inverse 
participation ratio. 

A large $\mbox{NPC}_j$ is associated with a delocalized state where many basis vectors 
give a significant contribution to the superposition $|\psi_j\rangle$;
whereas a small $\mbox{NPC}_j$ is related to
a localized state. Clearly, the components of the eigenvectors 
depend entirely on the choice of basis in which to express
them. Our approach here is to take a physically motivated basis
in order to study Anderson localization in disordered
systems with interacting particles.
Anderson localization refers to the 
exponential {\em spatial} localization of wavefunctions. 
This justifies our choice to consider
the site basis, which corresponds to a basis consisting of the eigenstates of $H_0$.
In this basis, 
the interaction $H_{\text{int}}$ contributes to the diagonal elements of the Hamiltonian, whereas
$H_{\text{hop}}$ constitutes the off-diagonal elements,
the latter being responsible for transfer of excitations
along the chain.

Because of its basis-dependence, NPC
is not an intrinsic indicator of quantum chaos.
To characterize the onset of chaos, we consider the 
level spacing distribution (see Sec.~II.D).
In studies of atomic and nuclear physics,
the basis-dependence of quantities to measure the complexity of wavefunctions
has long been pointed out~\cite{Zelevinsky1993}.
It has been argued that the mean-field basis is the preferred 
representation~\cite{Zelevinsky1993,Zelevinsky1996,ZelevinskyRep1996}, 
separating global properties from local fluctuations and correlations of the
wavefunctions.
Therefore 
in Sec.~III.A.2, we also briefly compare
the results for NPC in the site basis with those 
for two other representations:
free particles (FP)-basis and interacting particles (IP)-basis. The first 
consists of the eigenstates of $H_{\text{hop}}$ -- the integrable Hamiltonian 
describing a clean system
of free particles -- and the second 
consists of the eigenstates of $H_{\text{hop}}+H_{\text{int}}$ --
the Hamiltonian for a clean
chain with interacting particles. $H_{\text{hop}}+H_{\text{int}}$ 
is also an {\em integrable} model 
and is solved with the Bethe Ansatz method~\cite{Bethe}. 

Maximum delocalization, $\mbox{NPC} \sim N/3$, where $N$ is the matrix 
dimension~\cite{Guhr1998,Berman2001,Brown2008},
is obtained with
states from a chaotic system described by 
random matrices of a Gaussian Orthogonal Ensemble (GOE).
The hamiltonian studied here may also lead to chaos, but it 
is a banded random matrix, having only two-body interactions.
In this case,
the large values of NPC may be reached
only in the middle of the spectrum, 
the borders showing less delocalized states,
as typical of Two-Body Random Ensembles (TBRE) \cite{Brody1981,Kota2001}.
In TBRE's, the local density of states (LDOS) 
as a function of energy is Gaussian and peaked at the center
of the spectrum.

\subsection{Quantum Chaos}

In quantum systems, integrable and non-integrable regimes may be 
identified by analyzing the distribution of 
spacings $s$ between neighboring energy levels 
\cite{HaakeBook,Guhr1998}. 
Quantum levels of integrable systems tend to cluster and are not prohibited from crossing, in this case the 
typical distribution is Poissonian $P_{ P}(s) = \exp(-s)$.
In contrast, chaotic systems show levels that are correlated and crossings are strongly resisted, here
the level statistics is given by the Wigner-Dyson distribution. The exact form of the distribution
depends on the symmetry properties of the Hamiltonian. In the case of systems with time reversal invariance \cite{nota1}, 
it is given by $P_{ WD}(s) = \pi s/2\exp(-\pi s^2/4)$. 

For the purpose of illustration, instead of showing the level spacing distributions for the whole
range of parameters analyzed here, we compute $P(s)$ and associate with it
a number given by the quantity $\eta$, which is defined as

\begin{equation}
\eta \equiv \frac{\int_0^{s_0}[P(s) - P_{
WD}(s)]ds}{\int_0^{s_0}[P_{ P}(s)-P_{WD}(s)]ds},
\label{eta}
\end{equation}
where $s_0 \approx 0.4729$ is
the first intersection point of $P_P$ and $P_{WD}$. This quantity,
introduced in \cite{Jacquod1997}, simplifies the visualization of
the transition from integrability to chaos. For an integrable system: $\eta \rightarrow 1$, while 
for a chaotic system: 
$\eta \rightarrow 0$. In Ref.~\cite{Georgeot2000b}, 
$\eta = 0.3$ was taken as an arbitrary value below which the
system may be considered chaotic.
To derive meaningful level spacing distributions, 
besides unfolding the 
spectrum \cite{HaakeBook,Guhr1998}, all trivial symmetries of the system need to be
identified. 
The distributions are computed separately in each symmetry sector.

In addition to the conservation of $S^z$, the model described
by Eq.~(\ref{ham}) in the absence of disorder may also exhibit the following symmetries 
\cite{Brown2008}:  invariance under lattice reflection, which leads to parity conservation,
and, when the system is isotropic, 
conservation of total spin $S^2=(\sum_{n=1}^L \vec{S}_n)^2$, that is, $[H,S^2]=0$
(S$^2$ symmetry). Notice that only reflection symmetry exists in a chain with open boundary
conditions, whereas reflection and translational symmetries would occur in a ring. 

\subsection{Quantum correlations}

To quantify global quantum correlations, we consider the so-called global entanglement,
as proposed by Meyer and Wallach \cite{Meyer2002}.
This is a multi-partite entanglement measure employed for
lattices of two level systems (qubits).
For a pure state $|\psi_j\rangle$ of a chain with $L$ qubits, it is defined as

\begin{equation}
Q_j=2-\frac{2}{L}\sum_{n=1}^L Tr(\rho_n^2) ,
\end{equation}
where $\rho_n$ stands for the density matrix of the chain after tracing over all qubits
but $n$. $Q_j$ is therefore linearly related to the average purity of each two-level system, that is, it
is an average over the entanglements of each qubit with the rest of the system
\cite{Meyer2002,Brennen2003,Barnum2003}. Maximum global entanglement corresponds to $Q=1$.

Given the conservation  
of $S^z$ in $H$ (\ref{ham}) [of the number of particles in $H$ (\ref{ham_fermions})], 
this expression may be further simplified.

For spins: 
\begin{equation}
 Q_j=1-\frac{1}{L}\sum_{n=1}^L |\langle \psi_j|\sigma_n^z|\psi_j \rangle|^2.
\label{Q-spin}
\end{equation} 

For spinless fermions: 
\begin{equation}
Q_j=1-\frac{1}{L}\sum_{n=1}^L 
|\langle \psi_j|2 a^{\dagger}_n a_n - 1|\psi_j \rangle|^2.
\end{equation} 

A more general entanglement measure, the so called generalized entanglement (GE), 
was proposed in~\cite{Barnum2004}. GE is based 
on the relationship of the state with a distinguished set of observables, rather
than a distinguished subsystem decomposition. It coincides with $1-Q_j$
when the observable set corresponds to the expectation values of the
local magnetizations, $|\langle \psi_j|\sigma_n^{\beta}|\psi_j \rangle$ with
$\beta = x,y,z$.
Global entanglement is also closely related with the more broadly used von Neumann 
entropy~\cite{Brown2008}. All of the above are measures of multipartite entanglement, which differ
significantly from measures, such as concurrence~\cite{Wootters}, which aim at capturing
pairwise correlations.

The meaning of global entanglement is probably better understood with examples.
Consider, for instance, a bipartite system consisting of two qubits. Each qubit
may be in state $|0\rangle$ or $|1\rangle$. States such as
$|00\rangle$, $|01\rangle$, or even $[|00\rangle + |01\rangle + |10\rangle + |11\rangle]/2$
are not entangled, $Q=0$. The latter, in particular, is maximally delocalized, but not 
entangled, since it may be written as a product state 
$[|0\rangle+|1\rangle]/\sqrt{2}\otimes[|0\rangle+|1\rangle]/\sqrt{2}$.
In contrast, a state such as $a_1 |01\rangle+ a_2|10\rangle$, where
$|a_1|^2,|a_2|^2 \neq 0$, cannot be written as a product of states of the qubits; the 
qubits are then non-locally correlated and the state is entangled, $Q\neq0$. Maximum entanglement,
$Q=1$, occurs when $|a_1|^2=|a_2|^2 =1/2$, which corresponds to the so-called EPR or Bell state.
In this case, $Tr(\rho_n^2)=1/2$, that is, the reduced state of each qubit is maximally mixed.
But global entanglement goes beyond bipartite entanglement and quantifies the amount of 
multi-partite entanglement. The GHZ state, 
$[|00\ldots 00\rangle+|11 \ldots 11\rangle]/\sqrt{2}$,
or state $[|00\rangle+|11\rangle]/\sqrt{2}\otimes[|00\rangle+|11\rangle]/\sqrt{2}$
are examples of states with maximum 
global correlation~\cite{Brennen2003}.

\section{Half-filled chain}

We consider a half-filled one-dimensional system with $L=12$ and $M=6$.
This choice corresponds to the largest subspace of the
Hamiltonian, the sector where chaos sets in first.

\subsection{Uncorrelated random disorder}

\subsubsection{Site-basis}

\begin{figure}[htb]
\includegraphics[width=4.5in]{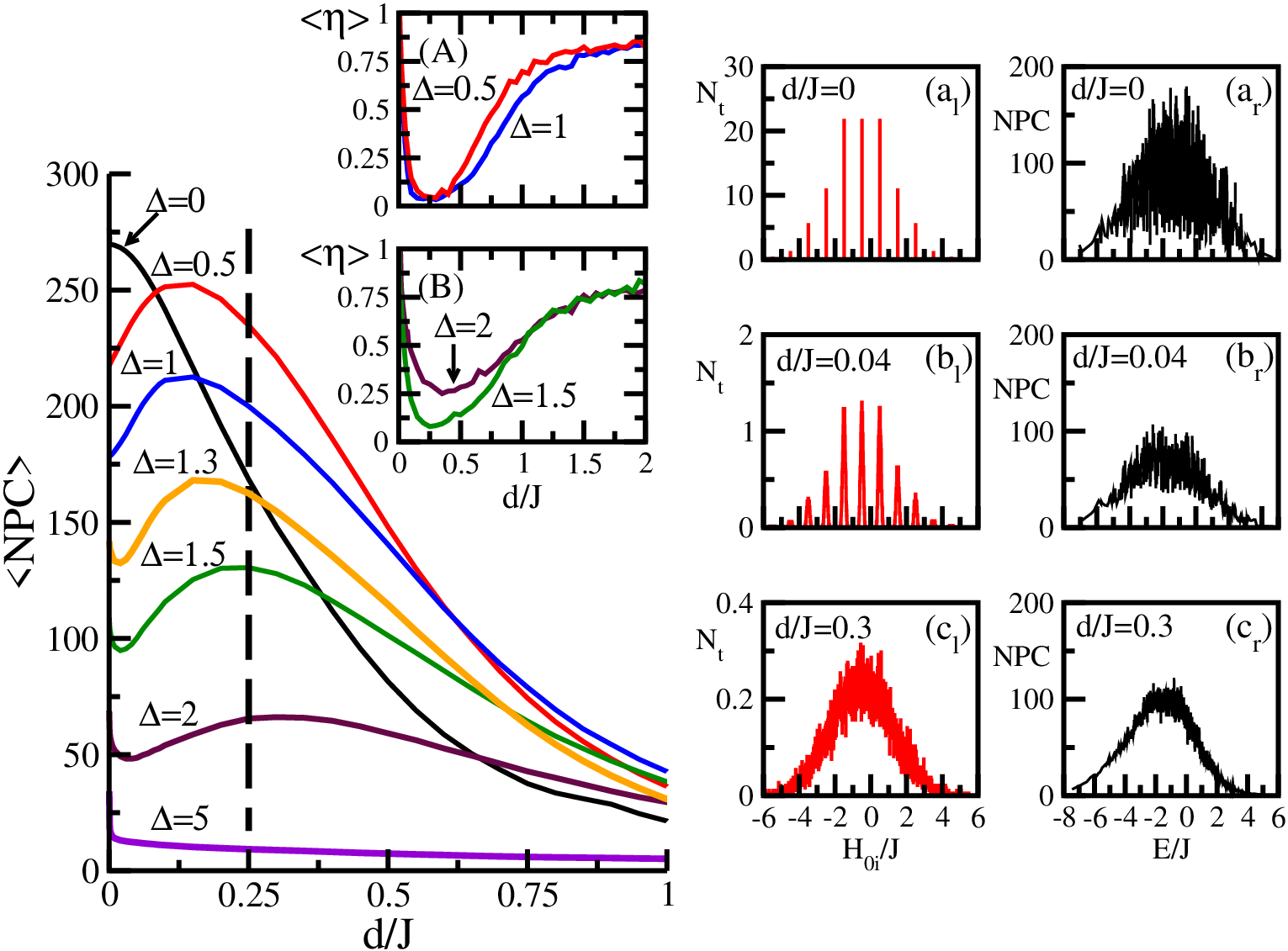}
\caption{Interplay between interaction and uncorrelated 
random disorder.
Main panel: NPC is the site-basis vs. $d/J$. 
Average over 924 states and 20 realizations.
Panels (A) and (B):  Level of chaoticity vs. $d/J$
(unfolding as described in ~\cite{Santos2004,notaCAPTION}); 
average over 20 realizations.
Panels (a$_{l,r}$), (b$_{l,r}$), (c$_{l,r}$): $\Delta =2$;
left: normalized histogram  
for the diagonal matrix elements $H_{0i}$ ($N_t$ is the number of states); 
right: $\mbox{NPC}$ vs. eigenvalues;
20 realizations.
All panels: $L=12$, $M=L/2$, Gaussian random numbers: $\alpha=0$.}
\label{fig1}
\end{figure}

We start by investigating how uncorrelated disorder affects the 
spatial localization of systems with interacting particles.
In the main panel of Fig.~\ref{fig1}, we plot the average NPC  in 
the site-basis vs. the amplitude $d/J$ of 
uncorrelated Gaussian disorder 
for different values of the interaction strength.
The maximum spatial delocalization occurs in a clean system in the absence of
interaction. 
At $d/J=0$, the level of delocalization decreases with $\Delta $,
and for $\Delta =0$, NPC also decreases with $d/J$. 
On the other hand, when both interaction and disorder are present, the behavior is 
not monotonic and NPC reaches a peak for $d/J<1$, eventually decreasing 
again as the disorder becomes larger than the 
hopping integral.

At $d/J=0$, the wavefunctions of the chain with interacting particles
are delocalized in the site-basis, although the system is integrable.
The addition of disorder in cases where $0<J\Delta \lesssim J$
further increases NPC by breaking symmetries
and allowing for couplings between more basis states.
This is completely antagonic to the
behavior of chains with non-interacting particles
or with very strong interactions ($J\Delta \gg J$), where disorder only localizes wavefunctions.
By comparing the curves of NPC with $\eta$ in 
panels A and B, one sees that
the delocalization peak 
is directly related to the onset of quantum chaos. 
There, the minimum value of $\eta$ 
for $\Delta$ = 0.5, 1 and 1.5 occurs at $d \sim J/4$.
However, it is important to emphasize that the bump in NPC should not be 
taken alone as an indication of chaos. In fact, 
a similar but less dramatic behavior is observed 
for the NPC of a two dimensional system, which is
chaotic already at $d/J=0$~\cite{Brown2008}.

To better understand the further delocalization
of wavefunctions in interacting 1D systems when disorder is considered,
we select the case where  $\Delta=2$
and compare, for three values of $d/J$,
the histograms of the diagonal elements $H_{0i}$ 
in panels a$_l$, b$_l$, c$_l$
and the plots of the level of delocalization of the eigenvectors of 
Hamiltonian (\ref{ham}, \ref{ham_fermions}) vs. 
their corresponding energies E in panels a$_r$, b$_r$, and c$_r$.

When $d/J=0$, the energies of the
basis vectors form narrow bands of 
resonant states (a$_l$), the energies being determined by the number of pairs of 
neighboring excitations
and by the number of excitations placed at the edges of the 
chain (border effects). These energies range from 
$-J \Delta (L-1)/4$ to $J\Delta (L-3)/4$ and the bands are separated by $J\Delta$.
As the hopping term is turned on,
only basis vectors belonging to the same symmetry sector can couple to form the wavefunctions.
In addition, if $J\Delta>J$, the effects of $J$ on states belonging to different bands 
are negligible and the bands, although now of finite width, remain separated.
In this case, only states placed in the same 
band and showing the same symmetries can mix, so the level of delocalization
is reduced to the number of states in the band. Moreover, the number of states
in a band decreases with the size of the clusters, that is, the 
number of neighboring excitations~\cite{Kagan1984}. The overall effect in clean systems is 
therefore the monotonic decrease of average NPC with interaction.
The right panel (a$_r$) shows the values of
NPC. For this integrable system, no clear relationship between
NPC and E is seen, as expected due to the absence of level repulsion.
It is interesting to contrast this behavior with the 2D clean system,
which is chaotic and
outlines the shape typical of chaotic TBRE~\cite{Kota2001} 
already at $d/J=0$ (data not shown).

By slightly increasing $d/J$, the bands broaden. If $J\Delta>J$, 
they remain uncoupled (b$_l$): 
the number of resonant intra-band states then decreases and so does NPC (b$_r$). 
This explains the valley that precedes the peak in the NPC curves
shown in the main panel.
Larger disorder is then needed to overlap the bands 
(c$_l$) and increase delocalization (c$_r$).
Notice that the average NPC for $d/J=0$ and 0.3 is approximately the same, 
although the NPC dependence 
on energy is significantly different (cf. a$_r$ and c$_r$). 
The integrable clean chain shows no relationship between NPC and E (a$_r$), but 
as complexity increases and $\eta$ decreases, a relationship 
similar to those appearing in TBRE's becomes evident (c$_r$):
delocalized states are in the middle of the spectrum and only 
localized states appear in the edges.

The case of $\Delta=2$ is at the borderline, 
where band overlapping is significant and where
level spacing distributions close to a Wigner-Dyson distribution may still be 
obtained. For larger interactions, such as $\Delta =5$, the bands are very separated
at $d/J=0$
and do not merge together by increasing $d/J$, instead, 
larger disorder simply prevents resonances leading to the monotonic decay
of NPC (see main panel).

\begin{figure}[htb]
\includegraphics[width=3.5in]{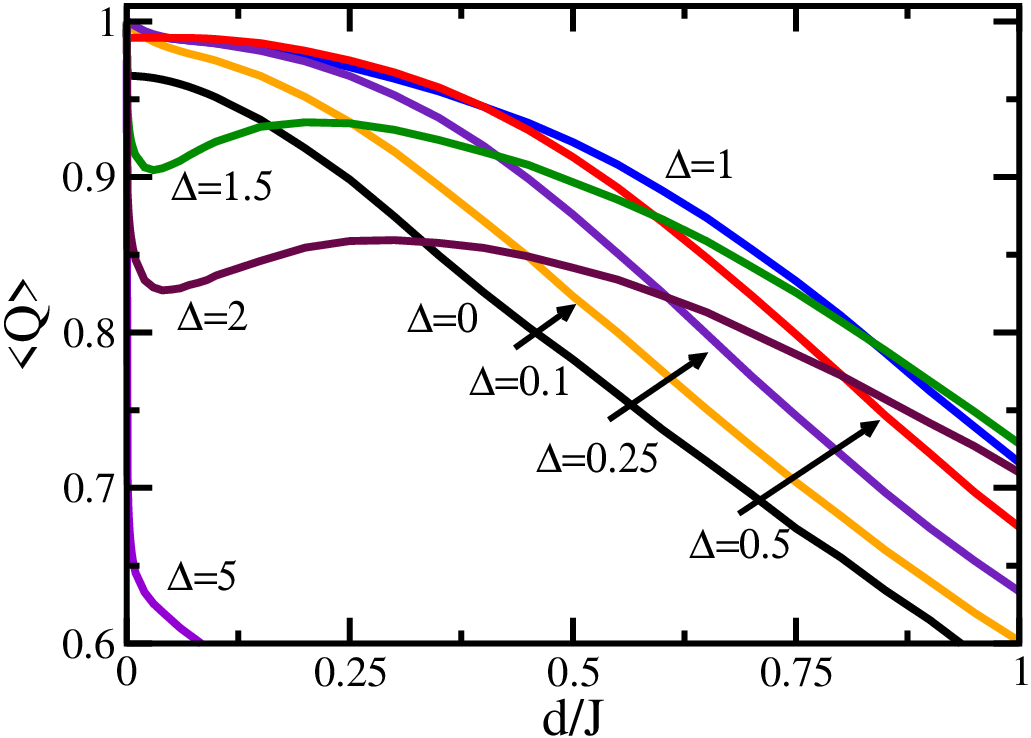}
\caption{Interplay between interaction and uncorrelated 
random disorder.
Average global entanglement vs. $d/J$. 
Average over 924 states and 20 realizations; $L=12$, $M=L/2$; $\alpha=0$.}
\label{fig2}
\end{figure}

Fig.~\ref{fig2} shows the behavior of global entanglement
versus uncorrelated disorder for various values of the
anisotropy \cite{nota3}.
A clean isotropic chain has $Q=1$, since the Hamiltonian is invariant under a global rotation of 180 degrees around the x-axis \cite{Brown2008}. For $d/J \rightarrow 0$, 
contrary to NPC, the largest value of $Q$ is found for a 
clean system in the presence of weak interaction, $0<\Delta<1$.
This suggests that interactions are key ingredients
for the generation of non-local correlations in a half-filled chain. 
Therefore, we identify two main factors
contributing for multipartite entanglement: the hopping term, which spreads the wavefunction
[a state with on-site localization obviously shows no
global entanglement], and the interaction term, which further enhances the
correlations between the particles.
High levels of entanglement require more than simply delocalization. As we discussed in 
Sec.~II.E, it is possible to have very delocalized states with low entanglement.

Also in contrast with NPC, when 
$\Delta \leq 1$, global entanglement decreases monotonically with disorder.
This is a result of the breaking of symmetries. It is possible, although this requires further investigation, that besides $S^2$ symmetry (and consequent rotational symmetry), reflection also has positive effects on quantum correlations. To illustrate this idea, compare the values of NPC and $Q$ in the table below. We have a system with $L=12$, $\Delta=0.5$, and two equal defects at the borders, $\omega_1=\omega_L=0.05J$, which guarantees that the $S^2$ symmetry is broken, but not parity. Suppose that we may include an additional defect on site $n$ with energy $0.5J$ (if $n=L$, then $\omega_L$ becomes $0.55J$). This is a simple way of breaking parity.

\begin{center}
\begin{tabular}{|l|l|l|}
\hline
& NPC & $Q$ \\
\hline
No additional defect  & 217 & 0.989 \\
\hline
Additional defect at $n=L$ & 176 & 0.953 \\
\hline
Additional defect at $n=(L/2)+1$ & 256 & 0.987\\
\hline
\end{tabular}
\end{center}
The chain with equal or unequal border defects is integrable~\cite{Alcaraz1987}, 
whereas a defect in the middle leads to chaos~\cite{Santos2004}. Independently of the position of the additional defect, it simply decreases $Q$. NPC, on the other 
hand, decreases only in the integrable regime, but increases in the chaotic region.
It has been argued in Ref.~\cite{Brown2008} that $Q$ seems closer related to
NPC than with the integrable-chaos transition. Here, we go one step further and claim that, even
though delocalization is necessary for the existence of quantum correlations, the two quantities
do not always go hand in hand. The onset of chaos has a positive effect on spatial delocalization,
whereas for global entanglement the breaking of symmetries associated with chaos
are more detrimental than possible gains associated with further delocalization.
%
%

Large interactions ($\Delta>1$)
are unfavorable to both NPC and $Q$, since the spectrum becomes gapped. 
However, for anisotropies not too large, $\Delta\gtrsim 1$, as $d/J$ increases and the energy bands 
overlap, the $Q$ curves quickly surpass the ones for smaller
interactions. This reinforces once again
the fundamental role of interactions 
in the creation of 
quantum correlations.

\subsubsection{FP- and IP-basis}

\begin{figure}[htb]
\includegraphics[width=4.5in]{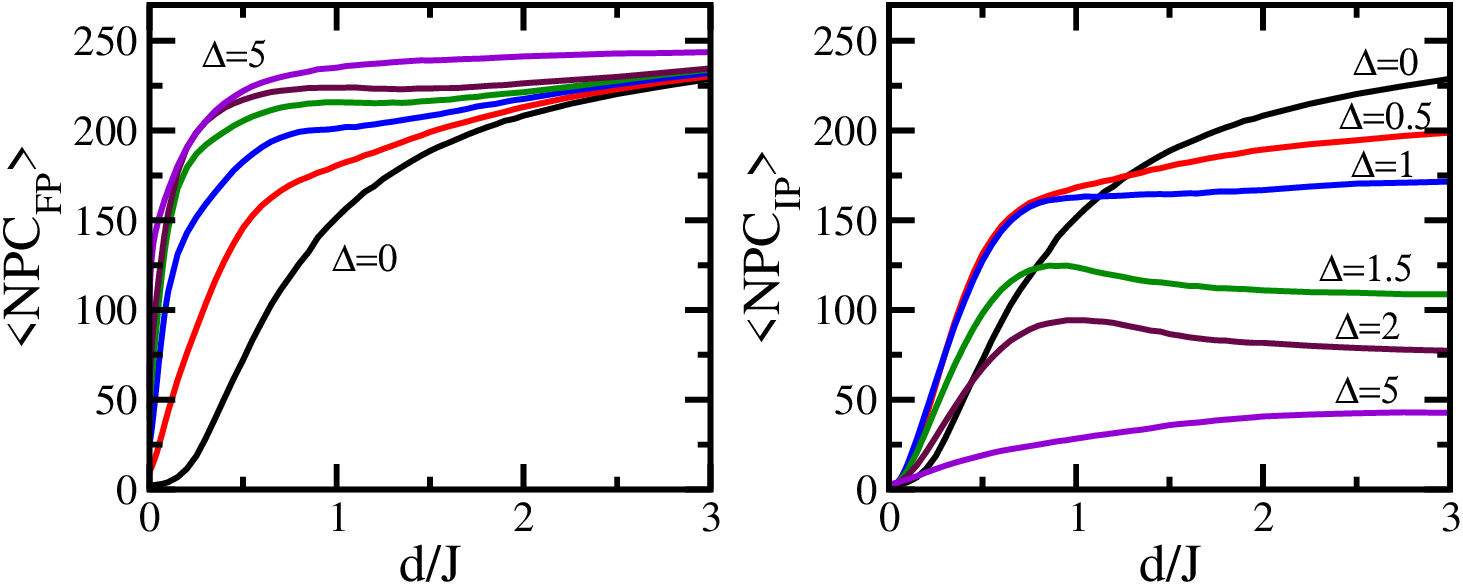}
\caption{ Average NPC vs. uncorrelated disorder.
Left: NPC in the free-particle basis;
from bottom to top: $\Delta = 0, 0.5, 1, 1.5, 2, 5$ Right: NPC in the basis
of the integrable interacting clean chain. 
Average over 924 states and 20 realizations; $L=12$, $M=L/2$; $\alpha=0$.}
\label{fig:NPC}
\end{figure}

The results for NPC are entirely dependent on the basis chosen
for the analysis. While the site basis is physically motivated
by studies of spatial localization, further insight may be gained
by considering alternative bases.
To separate the regular motion dictated by integrable Hamiltonians
from chaoticity, we show in Fig.~\ref{fig:NPC} 
the behavior of delocalization vs. disorder for NPC computed in 
the FP- and IP-basis.

The clean one-dimensional system considered here, with or without interactions, is integrable,
chaos emerges only when disorder is added to it.
In the FP-basis, a possible correlation between chaos and the complexity of the wavefunctions
is reflected in the rate of delocalization, which increases abruptly in the
chaotic region of $0<d/J<0.5$, $0<\Delta<1$, and is less dramatic for $\Delta=0$ and
$\Delta >2$. But overall, NPC simply increases with $\Delta$.
In contrast, in the IP-basis the behavior of NPC with $\Delta$ for $0<d/J<0.5$
is non-monotonic. Here, as in the site-basis case, delocalization is 
maximum where $\eta$ is minimum, the largest value appearing again for $\Delta=0.5$.
It is only at large values of disorder that the behavior of NPC with $\Delta$ becomes trivial.
Therefore, the effects of the interplay between interaction and disorder 
and the consequent onset of chaos are well singled out by the non-trivial behavior
of NPC
in both site- and IP-basis.

\subsection{Long-range correlated random disorder}

We now focus on a disordered chain with $d/J=0.25$ -- the 
chaotic region of interacting systems --
and proceed with a
comparison between uncorrelated and correlated random disorder.

\begin{figure}[htb]
\includegraphics[width=4.5in]{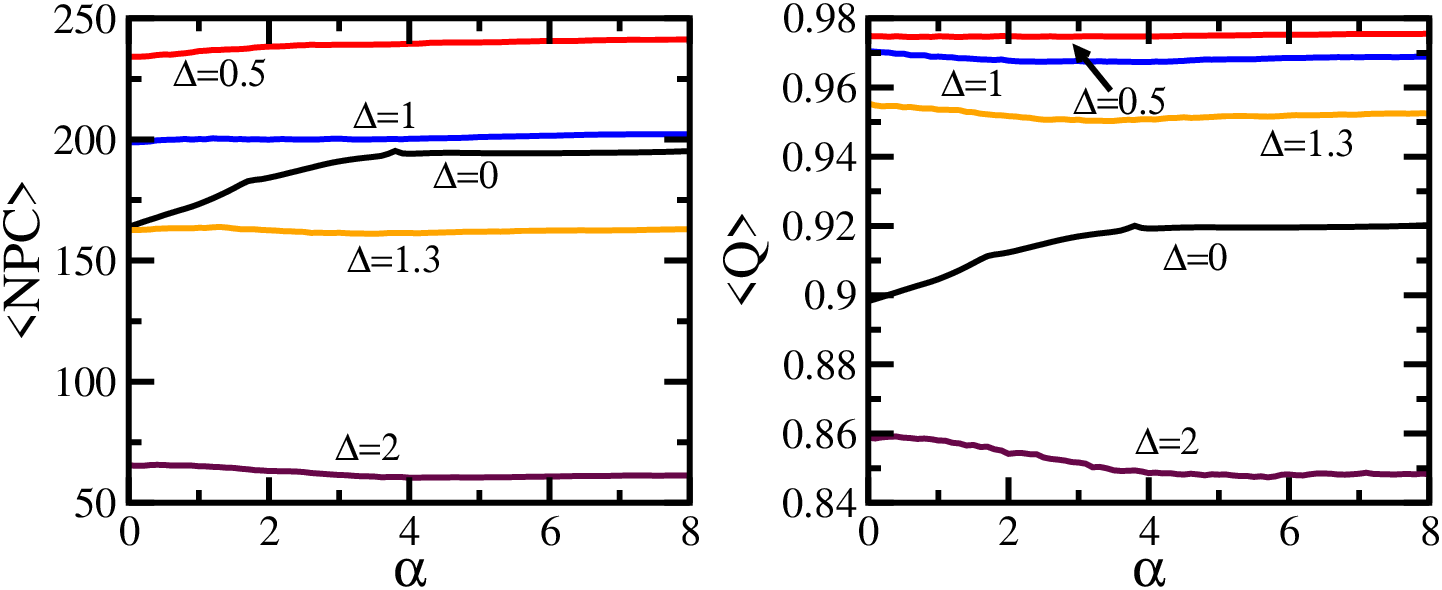}
\caption{Interplay between interaction and 
long-range correlated random disorder.
Left panel: Average delocalization in the site-basis vs. $\alpha$. 
Right panel: Average global entanglement vs. $\alpha$. 
All panels: $d/J=0.25$, $L=12$, $M=L/2$, average over 924 states and 20 realizations.}
\label{fig4}
\end{figure} 

The left panel of Fig.~\ref{fig4} shows the
average NPC vs. $\alpha$. 
The potential for spreading the wave functions
associated with $\alpha > 0$ has a non-monotonic behavior
with the anisotropy. 
This may be explained as follows. Correlated disorder indicates more order,
in a sense, it brings the system closer to a clean chain; but
how the system approaches its clean limit is highly
dependent on the interaction amplitude and the value of NPC we start with when $\alpha=0$.
Long-range correlated disorder can always increase delocalization when $\Delta=0$,
since a system with free particles is maximally delocalized in the 
absence of disorder (main panel in Fig.~\ref{fig1}). 
The growth in NPC is obviously not sufficient
to reach the values of a clean system, and not even those
obtained for chaotic systems with $\Delta < 1$, but, as shown in Fig.~\ref{fig4}, it 
approaches the level of delocalization of a chaotic chain with  $\Delta \sim 1$.
For interacting systems, on the other hand, correlated disorder may decrease NPC. 
The maximum value of NPC when $\alpha=0$ depends
on the amplitude of the uncorrelated disorder, $d/J$.
When $d/J=0.25$ and $\alpha=0$, it is seen from the main panel in 
Fig.~\ref{fig1} that the NPC curves with $\Delta<1$ have already passed their maximum, while those
with $\Delta\gtrsim 1$, such as $\Delta=2$, are right at their peak. Therefore, for $\Delta<1$,
by increasing $\alpha$ we can further delocalize the system, whereas
for $\Delta \gtrsim 1$, correlated disorder has the opposite effect of decreasing NPC.
For large values of $\Delta$, such as $\Delta=5$, 
since NPC simply decays with uncorrelated disorder (main panel of
Fig.~\ref{fig1}), correlated disorder will always increase delocalization.

\begin{figure}[htb]
\vskip 0.5 cm
\includegraphics[width=4.5in]{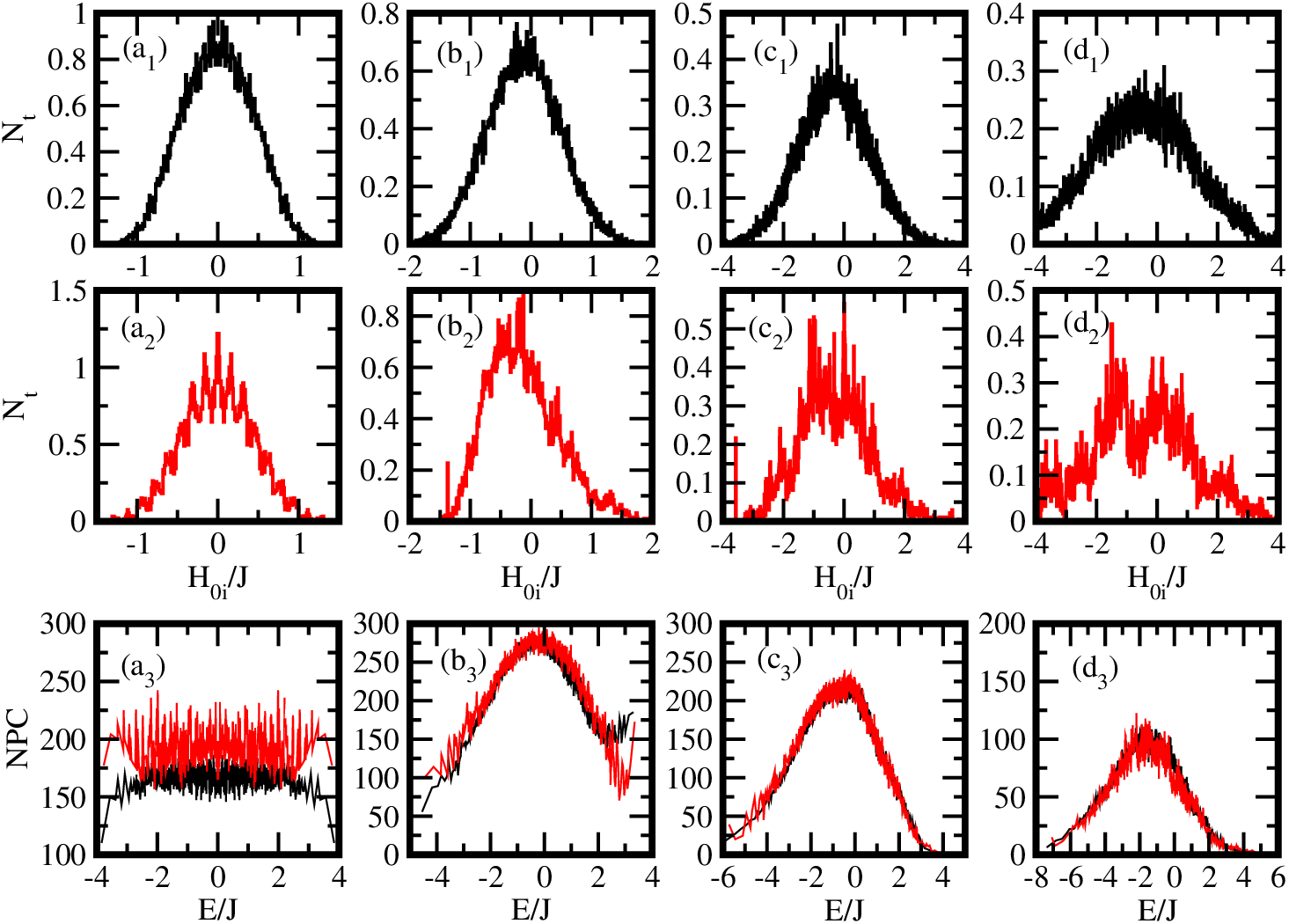}
\caption{Uncorrelated vs. long-range correlated random disorder.
First and second rows: normalized histograms for the diagonal matrix elements $H_{0i}$
with uncorrelated and correlated disorder, respectively.
Third row: $\mbox{NPC}$ vs. eigenvalues;
black curve: $\alpha=0$; red curve: $\alpha=10$.
Columns from left to right: $\Delta=0,0.5,1.3,2$.
All panels: $L=12$, $M=L/2$, 40 realizations.}
\label{fig3}
\end{figure}

Insight on the non-trivial behavior of NPC with $\alpha$
may be gained by
studying, in Fig.~\ref{fig3}, how correlated disorder affects
the histograms for the diagonal elements $H_{0i}$
and their consequences on the relation between NPC and the eigenvalues.
We select four values of the interaction amplitude, $\Delta$ = 0, 0.5, 1.3, and 2.0.

For uncorrelated disorder, 
as $\Delta$ increases, the distribution of $H_{0i}$ simply broadens
(Fig.~\ref{fig3} a$_1$, b$_1$, c$_1$, d$_1$),
however NPC does not decrease monotonically. 
As seen in the main panel Fig.~\ref{fig1}, for $d/J=0.25$,
$\Delta$ = 0 and 1.3 lead approximately to the same value
of NPC, delocalization is maximum at $\Delta \sim 0.5$
and is very low for $\Delta =2$.
This non-trivial behavior is associated with the transition
to chaos from $\Delta=0$ to $\Delta=0.5$: 
The featureless curve of NPC vs. E (a$_3$), 
which shows intermediate values of delocalization for all
energies, is substituted by a 
curve that peaks in the middle of the spectrum (b$_3$). 
As expected from chaotic TBRE's, for $\Delta=0.5$, NPC
approaches the GOE value of $N/3$ in the middle of the spectrum and 
has smaller values only at the edges. 
This results in an overall increase of $\langle$NPC$\rangle$.
It is only from $\Delta=0.5$ to $\Delta=1.3,2.0$,
that the broadening of the distributions of diagonal energies  
leads to the decrease of NPC (c$_3$,d$_3$). 

The inclusion of random disorder with long-range correlation ($\alpha=10$) has little effect 
on the delocalization of a half-filled system with 
interacting particles, but
it significantly affects the model with non-interacting particles.
When $\Delta=0$,  correlated 
disorder increases the number of resonances in the middle of the histogram
(compare a$_1$ and a$_2$), leading
to an interesting shaped 
distribution. This is reflected
in the larger values of NPC in panel a$_3$.
When $\Delta=0.5,1.3$, the shape of the distribution changes slightly
(cf. b$_1$, b$_2$ and c$_1$, c$_2$), having little effect on the curves 
for NPC vs. E (b$_3$, c$_3$). However, when $\Delta=2$ the energy band overlapping 
obtained by adding uncorrelated disorder (see a$_l$, b$_l$, and c$_l$ of
Fig.~\ref{fig1} and d$_1$ here) is partially removed by adding order
via $\alpha$ (d$_2$). This explains the decay of $\langle$NPC$\rangle$
verified in Fig.~\ref{fig4}. The effects of correlated disorder are, however, limited
and all curves for NPC and $Q$ saturate for $\alpha>4$
(see Fig.~\ref{fig4}).

There are then two possibilities to increase delocalization
in a chain of free particles with uncorrelated disorder: by including weak interaction,
so that chaos may set in, or by pushing the system toward the clean limit by adding 
correlated on-site energies.
The two alternatives 
constitute, however, opposite procedures, so when put together
their effects do not add up. 
%

In general, the behavior of multipartite entanglement 
and delocalization with correlated disorder are comparable. 
Similarly to NPC, the effects of $\alpha$ on $Q$ also depend
on the value of $Q$ we start with when $\alpha=0$, an information
extracted from Fig.~\ref{fig2}.
In the presence of correlated on-site energies,
the right panel of Fig.~\ref{fig4} shows that 
(i) global entanglement increases with $\alpha$ when $\Delta =0$,
(ii) $Q$ is not much affected by correlated disorder 
when $\Delta \sim 1$, (iii) $Q$ decreases with 
$\alpha$ when $\Delta \gtrsim 1$.
Notice, however, the
prominent role of interaction in establishing quantum correlations. 
In particular, compare the behavior of the curves for
$\Delta=0$ and 1.3. When $\alpha=0$, NPC$_{\Delta=0} \sim $ NPC$_{\Delta=1.3}$,
while $Q_{\Delta=1.3} \sim 1.06 \; Q_{\Delta=0}$. When 
$\alpha=8$, NPC$_{\Delta=0}$ increases significantly and 
becomes approximately 20\% larger than NPC$_{\Delta=1.3}$, while the growth of $Q_{\Delta=0}$ 
is more limited and it
remains smaller than the global entanglement for $\Delta=1.3$, 
$Q_{\Delta=1.3} \sim 1.03 \; Q_{\Delta=0}$.


\section{Dilute Limit}

The dilute limit implies $M\ll L$.
Here, we focus on the smallest value of $M$ where the interaction plays a role, the 
two particle case, $M=2$, and study the competition between interaction and 
both uncorrelated and correlated disorder.

\begin{figure}[htb]
\vskip 0.5 cm
\includegraphics[width=4.5in]{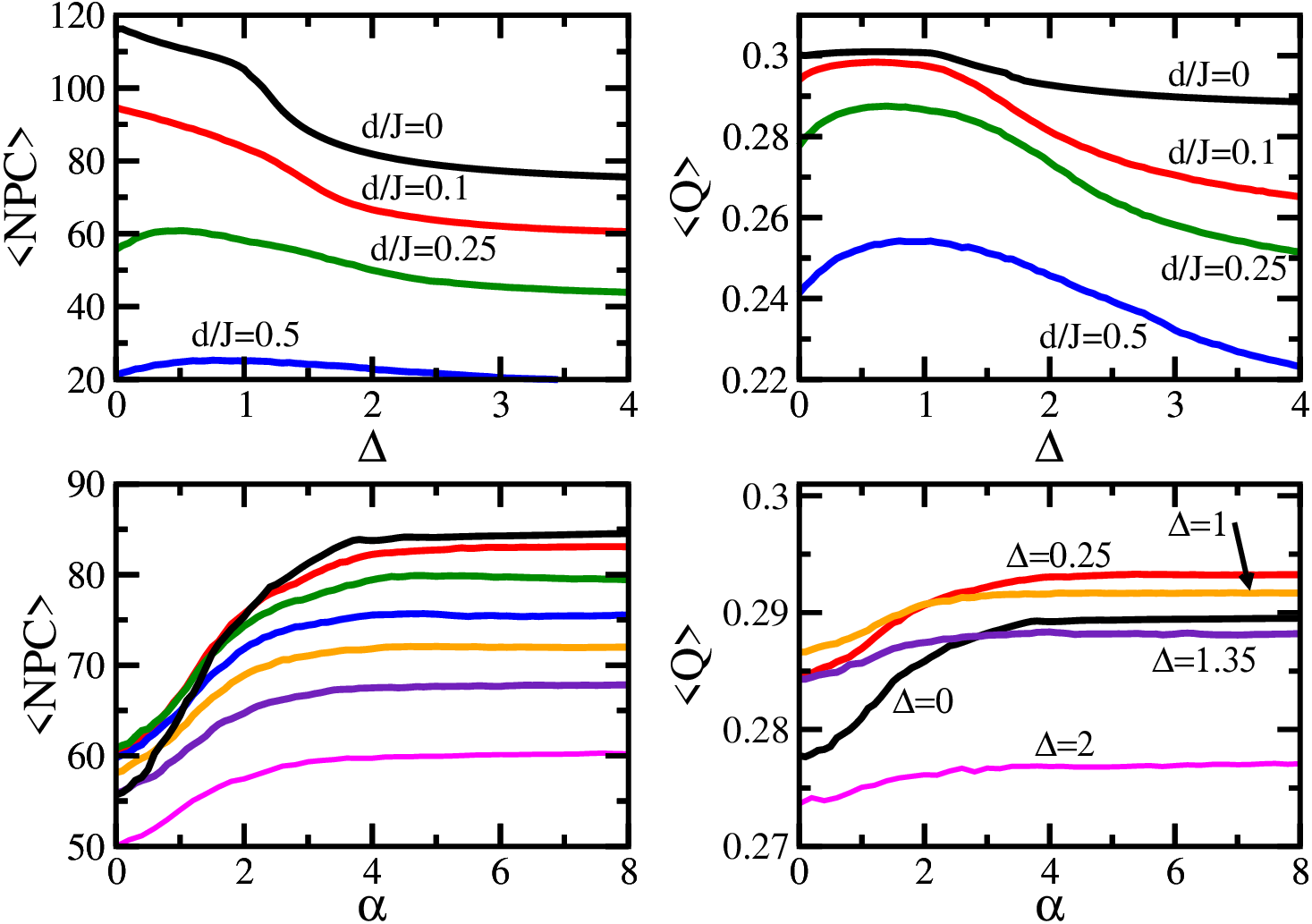}
\caption{Top panels: 
interplay between interaction and uncorrelated disorder, $\alpha=0$; bottom panels: 
interplay between interaction and long-range correlated disorder, $d/J=0.25$.
Top left (right) panel: Delocalization (Global entanglement) vs. anisotropy
for different values of $d/J$.
Bottom left (right) panel: Delocalization (Global entanglement) vs. $\alpha$
for different values of $\Delta$.
Left bottom panel - curves from top to bottom at
$\alpha = 8$: $\Delta =0,0.25, 0.5, 0.75, 1, 1.35, 2$.
All panels: $L=24$, $M=2$, average over all states and 20 realizations.}
\label{fig6}
\end{figure}

From the top panels of Fig.~\ref{fig6}, one sees that for a fixed value of the
interaction amplitude, NPC and $Q$ decay monotonically with disorder. 
This justifies our choice here
to consider $\Delta $ in the abscissa, instead of $d/J$, as in Figs.~\ref{fig1}, \ref{fig2}.
For $d/J\neq0$, such as $d/J=0.25$ and 0.5, a value of $\Delta\neq 0$ may still exist where NPC and $Q$
become maximum, but the effects of the interaction are now much less prominent
than in the half-filled chain.
In the dilute limit, the $H_{XY}$ term
dominates over the Ising interaction and becomes the main determining factor for
the level of delocalization and global entanglement. As a result, the behaviors 
of NPC and $Q$ are more comparable than at half-filling (cf. Fig.~\ref{fig6} 
and Figs.~\ref{fig1}, \ref{fig2}).

\begin{figure}[htb]
\hspace{-0.5cm}
\includegraphics[width=4.5in]{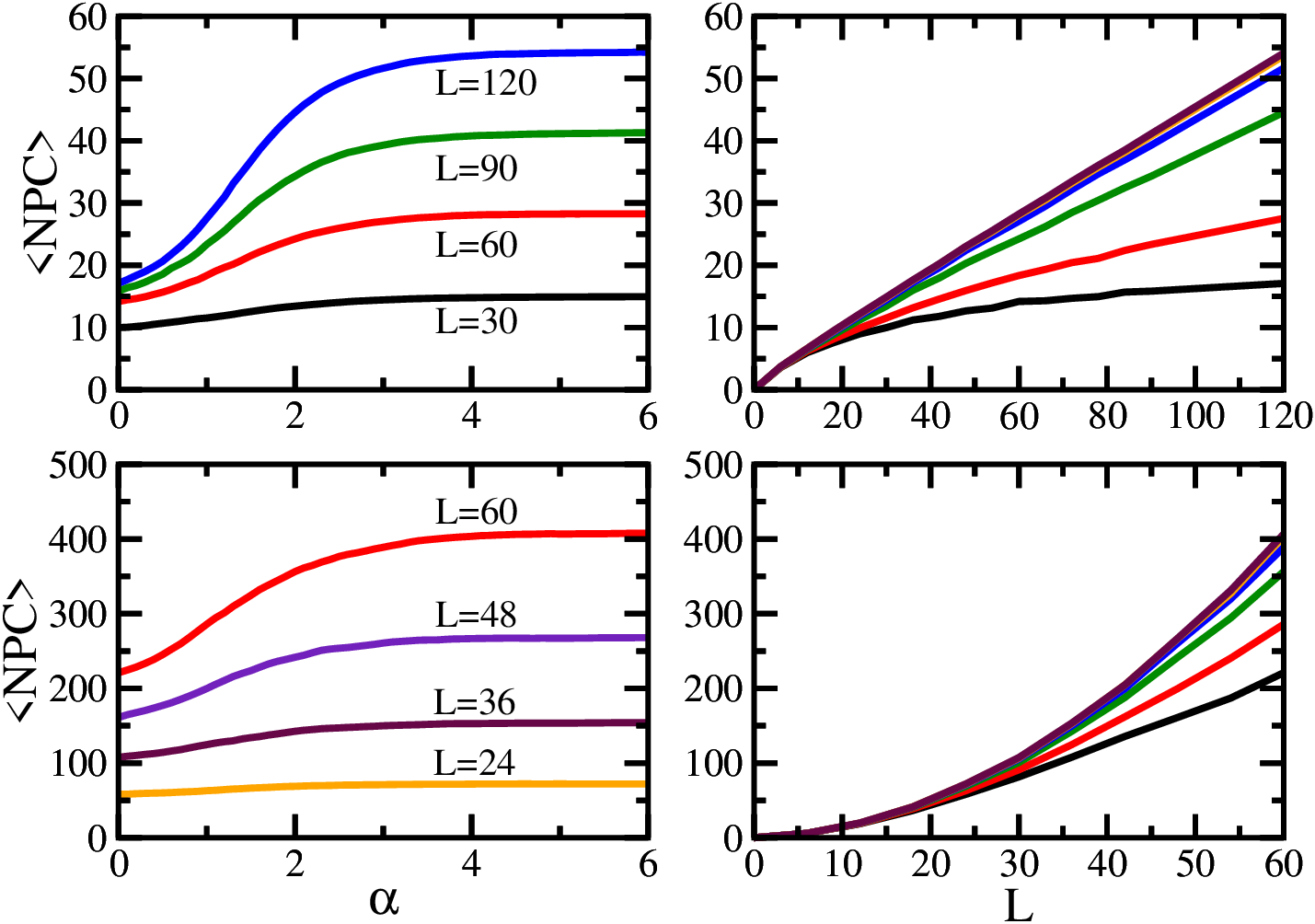}
\caption{Effects of the chain size 
on delocalization in the presence of long-range
correlated disorder. Top panels: $M=1$, bottom panels: $M=2$ and $\Delta=1$.
Left panels: Delocalization vs. $\alpha$ for different values of $L$; 
right panels: Delocalization vs. chain size for different values 
of $\alpha$, curves from bottom to top: $\alpha =0,1,2,3,4,5$. 
Average over all states and 40 realizations.}
\label{fig7}
\end{figure}

The bottom panels show how 
NPC and $Q$ behave with correlated disorder for various values of $\Delta$, keeping fixed $d/J=0.25$.
Contrary to the half-filled case, 
for all cases considered here, NPC and $Q$ increase with $\alpha$. This is a reflection
of the trivial behavior of NPC and $Q$ with uncorrelated disorder.
Moreover, the effects of $\alpha$ are now more pronounced than in the half-filled
chain, especially for the chain with $\Delta=0$. NPC$_{\Delta=0}$
now surpasses all curves shown and even
$Q_{\Delta=0}$ manages to outperform $Q_{\Delta=1.35}$. 
In comparison to the $M=L/2$ case, 
correlated disorder can now bring NPC and $Q$ closer to 
the values of a clean chain.
These results simply reflect the minor role of the interaction in a 
dilute system. The increase of NPC$_{\Delta=0}$
from $\alpha=0$ to $\alpha>4$ is more than 50\%
for $L=24$, $M=2$. As expected, this behavior
is yet amplified in larger systems, as shown in the left bottom panel
of Fig.~\ref{fig7}.

The right panels of Fig.~\ref{fig7} compare delocalization 
with system size in the case
of $M=1$ and $M=2$ for different values of $\alpha$.
When $M=1$ the dimension of the Hamiltonian is $N= L$.
For $\alpha=0$, as seen in the top right panel,
apart from small system sizes where border effects are important, there 
is little difference between the NPC values for different $L$'s.
This result is in agreement with the 
fact that Anderson localization occurs 
in one-dimensional systems with a single particle, that is,
uncorrelated random on-site energies lead to a finite
localization length of the wavefunctions. 
On the other hand, for large values of $\alpha$, the increase rate of NPC becomes linear. 
Large correlations between on-site energies bring
the behavior of NPC closer to that of a clean system, where NPC $\propto L$.
This is consistent with Ref.~\cite{Moura1998}, where it was shown that, in the
thermodynamic limit, one-particle states become delocalized for $\alpha>2$.

When $M=2$, the dimension of the Hamiltonian is quadratic on the system
size, $N=L(L-1)/2$. In this case, as shown in the right bottom panel,
NPC grows quadratically with $L$ even when $\alpha=0$.
This suggests that, on average and for the finite systems studied here,
there is no sign of Anderson localization for states with two interacting particles,
even when the disorder is uncorrelated. There are, however,
few isolated states with small NPC (data not shown), which accounts for situations where the 
particles are sufficiently apart, so that the interaction is negligible and 
the particles are individually localized. Further studies are now required
to conclude if, on average, localized states will be the dominating scenario
for larger chains, or if interaction will in fact prevent localization even
in the thermodynamic limit.

\section{Discussion and Conclusion}

We studied how delocalization and global entanglement in a 
quantum many-body one-dimensional system are
affected by the interplay between on-site (uncorrelated and long-range correlated) 
random disorder and nearest-neighbor interaction. 
The analysis applies equally to a chain with spins-1/2, spinless fermions, or
hard-core bosons. Half-filled and dilute chains were considered.

In a half-filled chain, the largest value of spatial 
delocalization  (NPC)
appears in a clean chain ($d/J =0$)
with non-interacting particles ($\Delta =0$). Here, the only constraints
in the spreading of the wavefunctions are the symmetries of the system
and the boundary conditions.   
The addition of uncorrelated disorder to a non-interacting system or
the inclusion of interaction  to a clean chain, suppresses resonances
and decreases NPC. 
On the other hand, when both 
$\Delta$ and $d/J$ are present,
the behavior of NPC becomes non-trivial.
A clean chain with interacting particles constitutes an integrable model
with eigenvectors delocalized in the site-basis. As uncorrelated
disorder is added, symmetries are broken. In the case
of  weak interaction, $0<\Delta \leq 1$, this leads to the onset of 
chaos and the further increase of NPC in the site-basis.

The behavior of global entanglement ($Q$) is influenced by delocalization, but 
does not follow it exactly. 
In particular, for $d/J \rightarrow 0$, the largest values appear in the presence of weak interaction. Also different,
in weakly interacting systems, is its 
steady decay with uncorrelated disorder and the consequent breaking of symmetries.
Besides delocalization, key contributing factors for global entanglement
are then interactions and symmetries. 

We have also briefly commented on the results for NPC in bases other
than the site-basis. The latter is considered in studies of spatial localization,
which was our main goal in this work.
However, it is also useful to separate regular 
motion from chaoticity by analyzing delocalization in a basis where the model
is integrable. In the basis consisting
of eigenstates of a clean system with interaction (IP-basis),
a non-monotonic behavior of NPC 
with $\Delta$ was also verified for disordered systems with $d/J < 1$.
This non-trivial behavior of NPC with interaction 
in both the IP- and the site-basis appears to be a main
indication of the complexity enhancement of 
wavefunctions in the chaotic region.

The presence of correlated disorder ($\alpha \neq 0$) has different 
consequences depending on 
the value of $\Delta$. By correlating on-site energies, 
the level of order increases.
A disordered system with non-interacting particles then
gets closer to the limit of a clean chain and NPC increases.
In the presence of interaction, the effect is opposite, since here
it is disorder that enhances delocalization by breaking symmetries and 
overlapping energy bands. The behaviors of NPC and $Q$ with $\alpha$
reflect the non-trivial behavior of these quantities with $d/J$ when $\alpha=0$.

The study of a dilute system allows for 
the consideration of larger chains. In the presence of 
uncorrelated disorder, contrary to the one-particle case, on average, our results
show no sign of Anderson localization in the finite systems considered.
Long-range correlation delocalize both one and two-particle states.

In a future work, we intend to search for conclusive evidences for the absence (or not) of
localization of two-interacting-particle states in systems with uncorrelated disorder.
We also aim to study in details the effects of symmetries on non-local correlations.
Another question worth investigation is how
the integrable and chaotic regimes of the considered one-dimensional systems 
relate to different transport behaviors, namely diffusive or ballistic~\cite{Santos2008PRE},
and to the problem of thermalization~\cite{ZelevinskyRep1996,Rigol2008}.
Also interesting would be the 
comparison of results with the Hubbard model and the
inclusion of long-range interactions to the systems considered.

We hope that our theoretical results will motivate experimental verifications. 
Useful tools to simulate many-body effects in simplified models of 
condensed matter physics as the ones described here are
optical lattices. They 
allow for control of the parameters of the system, such as
the strength of interaction and the level of disorder~\cite{Greiner2008}.

\begin{acknowledgments}
M.Z. thanks Stern College for Women at Yeshiva University for a summer fellowship during the 
initial development of this project. This work was supported by the Research Corporation.
\end{acknowledgments}


\begin{thebibliography}{65}
\bibitem[{not({\natexlab{b}})}]{notaCAPTION}
\bibinfo{note}{At $d/J = 0$, the correct evaluation of $\eta$ would need to
  take into account the symmetries of parity and, when $\Delta=1$, total spin; moreover, when
  $J\Delta>J$, a specific energy band, determined by the number of pairs of
  excitations and border excitations, would also need to be selected.}
\end{thebibliography}
\end{document}